\input harvmac

\def\Title#1#2{\rightline{#1}\ifx\answ\bigans\nopagenumbers\pageno0\vskip1in
\else\pageno1\vskip.8in\fi \centerline{\titlefont #2}\vskip .5in}
%

%
%
\ifx\includegraphics\UnDeFiNeD\message{(NO graphicx.tex, FIGURES WILL BE IGNORED)}
\def\figin#1{\vskip2in}
\else\message{(FIGURES WILL BE INCLUDED)}\def\figin#1{#1}
\fi
\def\Fig#1{Fig.~\the\figno\xdef#1{Fig.~\the\figno}\global\advance\figno
 by1}
%
%
%
%

%
%
\font\ticp=cmcsc10
\font\bbbi=msbm10
\def\bbb#1{\hbox{\bbbi #1}}

\def\calo{{\cal O}}
\def\calh{{\cal H}}
%
%
\lref\GiddingsPJ{
  S.~B.~Giddings,
  ``Black holes, information, and locality,''
  arXiv:0705.2197 [hep-th].
}
\lref\Hartone{
  J.~B.~Hartle,
  ``The Quantum mechanics of cosmology,''
  in {\sl Quantum cosmology and baby universes : proceedings}, 7th Jerusalem Winter School for Theoretical Physics, Jerusalem, Israel, December 1989, ed. S. Coleman, J. Hartle, T. Piran, and S. Weinberg (World Scientific, 1991).
}
\lref\Harttwo{
  J.~B.~Hartle,
  ``Space-time coarse grainings in nonrelativistic quantum mechanics,''
  Phys.\ Rev.\  D {\bf 44}, 3173 (1991).
}
\lref\HartLH{
  J.~B.~Hartle,
  ``Space-Time Quantum Mechanics And The Quantum Mechanics Of Space-Time,''
  arXiv:gr-qc/9304006.
}
\lref\HartPuri{
  J.~B.~Hartle,
  ``Quantum Mechanics At The Planck Scale,''
  arXiv:gr-qc/9508023.
}
\lref\HartleNX{
  J.~B.~Hartle,
  ``Generalizing quantum mechanics for quantum spacetime,''
  arXiv:gr-qc/0602013.
}
\lref\Wein{
  S.~Weinberg,
 ``The Quantum theory of fields. Vol. 1: Foundations,''
{\it  Cambridge, UK: Univ. Pr. (1995).}
}
\lref\GiMa{
  S.~B.~Giddings and D.~Marolf,
  ``A global picture of quantum de Sitter space,''
  arXiv:0705.1178 [hep-th].
}
\lref\Eden{R.J. Eden et. al., {\sl The analytic S-Matrix} (1966, Cambridge University Press).}
\lref\GiLi{
  S.~B.~Giddings and M.~Lippert,
  ``Precursors, black holes, and a locality bound,''
  Phys.\ Rev.\  D {\bf 65}, 024006 (2002)
  [arXiv:hep-th/0103231].
}
\lref\LQGST{
  S.~B.~Giddings,
  ``Locality in quantum gravity and string theory,''
  Phys.\ Rev.\  D {\bf 74}, 106006 (2006)
  [arXiv:hep-th/0604072].
}
\lref\GMH{
  S.~B.~Giddings, D.~Marolf and J.~B.~Hartle,
  ``Observables in effective gravity,''
  Phys.\ Rev.\  D {\bf 74}, 064018 (2006)
  [arXiv:hep-th/0512200].
}
\lref\GaGi{
  M.~Gary and S.~B.~Giddings,
  ``Relational observables in 2d quantum gravity,''
  arXiv:hep-th/0612191, Phys.\ Rev.\  D {\bf 75} 104007 (2007).
}
\lref\AiSe{
P.~C.~Aichelburg and R.~U.~Sexl,
``On The Gravitational Field Of A Massless Particle,''
Gen.\ Rel.\ Grav.\  {\bf 2}, 303 (1971).
}
\lref\GiddingsBE{
  S.~B.~Giddings,
 ``(Non)perturbative gravity, nonlocality, and nice slices,''
  Phys.\ Rev.\  D {\bf 74}, 106009 (2006)
  [arXiv:hep-th/0606146].
}
\lref\GiddingsSJ{
  S.~B.~Giddings,
  ``Black hole information, unitarity, and nonlocality,''
  Phys.\ Rev.\  D {\bf 74}, 106005 (2006)
  [arXiv:hep-th/0605196].
}
\lref\HartleAI{
  J.~B.~Hartle and S.~W.~Hawking,
 ``Wave Function Of The Universe,''
  Phys.\ Rev.\  D {\bf 28}, 2960 (1983).
  }
  \lref\Dewitt{
B. DeWitt, ``The Quantization of Geometry'', in
{\it Gravitation: An Introduction to Current Research},  ed. Witten L (New
York, Wiley, 1962).}
\lref\Rove{
  C.~Rovelli,
  {\sl Quantum Gravity}, 
{\it  Cambridge, UK: Univ. Pr. (2004).}
}
\Title{\vbox{\baselineskip12pt
}}
{\vbox{
\centerline{Universal quantum mechanics}
}}
\centerline{{\ticp Steven B. Giddings\footnote{$^\ast$}{Email address: giddings@physics.ucsb.edu} } }
\centerline{\sl Department of Physics}
\centerline{\sl University of California}
\centerline{\sl Santa Barbara, CA 93106}
\bigskip
\centerline{\bf Abstract}
If gravity respects quantum mechanics, it is important to identify the essential postulates of a quantum framework capable of incorporating gravitational phenomena.  Such a construct likely requires elimination or modification of some of the ``standard" postulates of quantum mechanics, in particular those involving time and measurement.  This paper proposes a framework that appears sufficiently general to incorporate some expected features of quantum gravity.   These include the statement that space and time may only emerge approximately and relationally.  One perspective on such a framework is as a sort of generalization of the S-matrix approach to dynamics.  Within this framework, more dynamical structure is required to fully specify a theory; this structure is expected to lack some of the elements of local quantum field theory.    Some aspects of this structure are discussed, both in the context of scattering of perturbations about a flat background, and in the context of cosmology.

\Date{}

\newsec{Introduction}

There are significant indications that formulation of a physical theory properly incorporating both quantum phenomena and gravity goes beyond the framework of local quantum field theory.\foot{For a brief overview of locality issues in gravity, see \GiddingsPJ.}  An important question is what basic physical and mathematical principles should be part of a theory, and conversely, what aspects of existing local quantum field theories should be abandoned.  Local quantum field theory is based on the principles of 1) quantum mechanics 2) Lorentz invariance and 3) locality.  There are different possibilities for which of these should be dropped, which have been considered in investigations over the years.  

In particular, it is certainly possible that we must abandon quantum mechanics.  However, numerous experimental verifications of quantum phenomena indicate that they are a part of nature.  Moreover, while one might imagine that there could be small violations of quantum mechanics, quantum theory has been remarkably robust, and has in particular proven difficult to consistently modify without serious conflict with other basic physical principles and/or observations such as energy conservation, stability, and causality.  Thus this note will adopt the viewpoint that quantum mechanics is an essential aspect of physical theory.

If quantum mechanics is indeed essential in physics, it is important to understand what is the appropriate structure needed in a quantum theory that can be reconciled with gravity.
One answer to this has been proposed in the form of {\it generalized quantum mechanics}\refs{\Hartone\Harttwo\HartLH-\HartPuri}.  Central in that framework is the existence of {\it histories}\foot{Or, more generally, {\it configurations} or 
{\it descriptions}\refs{\HartleNX}.} (both fine- and coarse-grained).  However, one theme in some of the thinking on the nature of quantum gravity is that space and time may not be fundamental, but instead may only emerge in an approximate and relational fashion in certain states and in the description of certain observables, as  for example explained in \refs{\GMH}.  This in turn suggests the possibility that the notion of a {\it history} may not be a central feature of the postulates.  

This note will investigate the question of what logical structure might be needed to describe such a quantum theory of physics.  The framework   appears more general than that of \refs{\Hartone\Harttwo\HartLH-\HartPuri}, and will be referred to by the name ``universal quantum mechanics."
The necessary postulates will be remarkably sparse.  In particular they do not include a fundamental notion of time, or other spacetime structure, which may only appear (semi)classically in certain states of certain theories.

This discussion first outlines a set of basic postulates.  To add context, it then briefly reviews the structure that quantum field theory adds to these postulates.  Finally, the discussion turns to the question of what structure might be expected in a quantum theory incorporating gravity, both in describing  gravitational scattering in ``asymptotically Minkowski space," and in describing cosmologies such as de Sitter space.

We will see that the postulates of this framework are similar to some of the postulates of the S-matrix program.  However, as we'll find, the current framework generalizes that program in several directions.

\newsec{Universal quantum mechanics}

\subsec{Postulates}

Essential features of quantum theory are the notions of states, linearity, and an inner product.  Thus basic postulates of universal quantum mechanics (UQM) are

{\it Postulate 1.}  {\sl Physical configurations are described by states  $|\psi\rangle$ (or more precisely, rays) of a vector space $\calh$, which is a vector space over the complex numbers.  A basis of this space is denoted $|\psi_\alpha\rangle$, where the parameters $\alpha,\beta,\ldots$ are elements of a set $A$ of labels.  This set may be finite or infinite, discrete or continuous (or both).  }

{\it Postulate 2.} {\sl The space $\calh$ has an inner product, linear under addition and scalar multiplication, in the familiar fashion.  The inner product of two states $|\psi\rangle$ and $|\psi'\rangle$ is denoted $\langle \psi|\psi'\rangle$.  }

As usual,  the inner product also induces a map from  $\calh$ to  the dual space $\calh^*$, mapping a state $|\psi\rangle$ to $\langle\psi|$.

For some of the systems we will consider, it proves useful to introduce a third, possibly optional, postulate:

{\it Postulate 3.} {\sl  In addition to the space $\calh$, there exists a complex vector space $\calh^d$, called a ``descriptor space"  or ``extended space," which is taken to contain $\calh$. 
Each state $|\varphi\rangle_d$ in the descriptor space gives a linear functional on $\calh$, written $|\psi\rangle\rightarrow {}_d\langle\varphi|\psi\rangle$.  
}

A space like  $\calh$, with a positive inner product, is what is sometimes called a ``pre-Hilbert space." Separability will not necessarily be assumed.  In the case of a true Hilbert space (satisfying completeness),
 the dual space $\calh^*$ is isomorphic to $\calh$.  The third optional postulate introduces the possibility that  there may be  a useful notion of a space $\calh^d$ that is bigger than the space of physical states. An example where such a space may be useful is in de Sitter cosmology: see section five.
 (We will not commit to this space being a  Hilbert space.)  
 A basis of this space is denoted $|\varphi_i\rangle_d$, where indices $i,j,\ldots$ are taken to lie in a set $A^d$.

We will also, of course, need to consider linear (and in particular hermitian) operators acting on $\calh$, and possibly on $\calh^d$.  These can for example be written in the form 
\eqn\opdef{ \calo= \sum_{\alpha,\beta\in A}\calo_{\alpha\beta} |\psi_\alpha\rangle\langle\psi_\beta|\quad ,\quad \calo^d= \sum_{i,j\in A^d}\calo_{ij} |\varphi_i\rangle_d{}_d\langle\varphi_j|\ .}
(Note that the latter operator has image in $\calh^d$.)

\subsec{Interpretation}

We next briefly outline an approach to interpreting the structure described in the preceding subsection.  First, the states in $\calh$ are thought of as the ``allowed physical states."  Then, if  states $|\psi\rangle$ and $|\psi'\rangle$ are unit norm, the complex number $\langle \psi' | \psi\rangle$ is interpreted as the probability amplitude for the state $\psi$ to ``take on the appearance of" (or ``look like") the state $\psi'$.  For example, in the case where a spinning particle is interacting with a detector that can measure the spin to be up or down, at some time $t$ specified by a dynamical clock, we have amplitudes for the detector at time $t+\delta t$ to register spin up, or spin down.  In the usual Copenhagen interpretation, there is a corresponding projection onto alternatives.  However, in the present framework both alternatives may be present in the wavefunction, which may thus ``take on the appearance of" either alternative, as in the relative-state or many-worlds formulation.

There is no intrinsic notion of time or history in this description.  Such notions may emerge for certain UQM theories in certain states.  For example, the state $\psi'$ may be a state of the Universe that describes two particles that collide in a detector when other macroscopic variables effectively defining ``time" (such as macroscopic clocks, positions of planets, stars, and galaxies) reach certain values.  In other words, time, in cases where it is defined,  {\it may only enter relationally in a particular kind of specification of the state.  }

One example of such a framework that illustrates the inessential nature of ``time" and ``history" is the original S-matrix program.  There one gives  a set of labels for what are called ``in" and ``out" states (vacuum, one-particle, two-particle, etc.).  The full content of the theory is then in the S-matrix elements between such states.  In this story, there is no a-priori reason that computation and interpretation of the S-matrix has to come from hamiltonian evolution in time, or a sum over histories.  One does, however, need the additional structure provided by the labeling of the in and out states in order to interpret the amplitudes.  But this labeling is part of the specification of the {\it particular dynamical system} and thus this  labeling is not part of the general quantum-mechanical framework. Indeed,  the specifics of the labeling (vacuum, one-particle state, two-particle states,  etc.) depend on the spectrum, which in turn depends on the dynamics.
As will be described, in more general contexts (e.g. cosmology) an interpretation relevant to the specific dynamical system should emerge from such a labeling, which likewise is expected to be given as part of the specification of the dynamics.

In some contexts it appears useful to broaden the class of states that we consider beyond those that are ``physical."  For example, one may wish to study amplitudes for a physical state to have the appearance of a state that is not truly a physical state.  Examples will be discussed in the cosmological setting.  For example, in the case of de Sitter space, one may consider amplitudes of de Sitter states to appear as states that are not legitimate physical states of global de Sitter space\refs{\GiMa}, as will be described.  This is the reason for introducing the descriptor space $\calh^d$, and in this context ${}_d\langle\varphi|\psi\rangle$ has an interpretation precisely paralleling the above.  Given the projection $\Pi:\calh^d\rightarrow\calh$, in cases where $\calh$ is Hilbert, such inner products reduce to those with states in $\cal H$, but the additional structure in ${\cal H}^d$ may prove useful nonetheless.

One can also discuss, and may find useful, expectation values of operators, or strings of operators, as well as more general matrix elements.  Moreover, as in usual quantum mechanics, states may be characterized by the action of these operators; {\it e.g.} by eigenvalues of certain operators.  For example, as in usual quantum mechanics, we may choose to work in a basis that diagonalizes a given operator $\calo$,
\eqn\diagop{\calo = \sum_{\lambda,a} \lambda | \lambda, a\rangle\langle \lambda, a|\ ,}
where $\lambda$ is an eigenvalue, and $a$ represents additional labels necessary to give a complete basis.  (More generally, one may wish to work with a set of commuting operators such that their collection of eigenvalues completely label the basis.)  One may thus take as a starting point for labeling states specification of a preferred set of such operators, or of a preferred basis of states, with
equivalence of these two approaches following from an expression like \diagop.

\subsec{Relation to generalized quantum mechanics}

A previous proposal for generalizing text-book quantum mechanics to a framework appropriate for describing gravity is ``generalized quantum mechanics" (GQM), introduced in \refs{\Hartone\Harttwo\HartLH-\HartPuri}.\foot{A related framework is Rovelli's ``relativistic quantum mechanics\refs{\Rove}."} Basic ingredients of GQM are 1) the sets of fine-grained histories; 2) a notion of coarse graining; and 3) a measure called the decoherence functional which establishes when quantum interference between coarse-grained histories is sufficiently negligible to admit a probabilistic interpretation.  While it is not clear that it is the only way generalized quantum theory can be applied, in all cases where it has been worked out so far a precise and local description of the fine-grained histories has been assumed. For example these are field configurations in the case of field theory and spacetime geometries in the case of general relativity.

However, it is not a-priori obvious that the notion of ``history" is an essential element in determining and describing amplitudes of a quantum system governing gravity.  Even in the simpler context of the S-matrix, the notion of a history (which, for an S-matrix derived from quantum field theory, corresponds to a classical field configuration) is not an a-priori necessary ingredient, and one might imagine existence of consistent S-matrix theories that do not emerge from a sum over such histories.  In gravity, this perspective  is reinforced by the expectation that there is not a precise notion of space, time, and locality.  Thus UQM is expected to be a more general framework than GQM, relevant in cases where ingredients 1)-3) are not available, or precise, or relevant.

Since UQM represents a generalization of GQM, we do expect UQM to reproduce the results of GQM when there is a notion of history available.
Specifically,  in certain UQM theories, there may be some preferred set of basis states, $|\psi_\alpha,t\rangle$, for each $t$, where $t$ is a real parameter ranging over some interval $[T_i,T_f]$.  In that case, we might define a ``history" in terms of a product of the form:
\eqn\histdef{\prod_t |\psi_{\alpha_t},t\rangle\langle \psi_{\alpha_t},t|\ ,}
which depends on a choice of a definite $\alpha_t$ for each $t$.  More precisely, the expression \histdef\ is a class operator\refs{\Hartone\Harttwo\HartLH-\HartPuri}, whose action on a state $|\psi\rangle$ gives the final state of the history, $|\psi_{\alpha_{T_f}}, T_f\rangle$, times the amplitude to follow the particular history from $|\psi\rangle$ to that final state.  (A discretization may useful in carefully defining such expressions.)\foot{It may also provide useful to generalize this definition of history to include states in $\calh^d$.}

To match the terminology of \refs{\Hartone-\HartPuri}, more precisely \histdef\ corresponds to a {\it fine-grained} history.  Class operators corresponding to {\it coarse-grained} histories may be defined by summing over fine-grained histories, for example over a range of $\alpha_t$ for each $t$.

Likewise, one can define the notion of decoherence functional.  Let $\{C_A\}$ be a set of such coarse-grained histories $C_A$, labeled by $A$.  Given a density matrix formed from states in $\calh$,
\eqn\densmat{\rho = \sum_{\alpha\beta} \rho_{\alpha\beta} |\psi_\alpha\rangle\langle\psi_\beta|\ ,}
one writes the decoherence functional, as in \refs{\Hartone\Harttwo\HartLH-\HartPuri},
\eqn\decohf{D(A',A) = {\rm Tr}\left(C_{A'} \rho C_A\right)\ .}
Then, the set of coarse grained histories are said to decohere to the extent this becomes approximately diagonal:
\eqn\decoh{D(A',A) \approx p_A \delta_{A'A}\ .}

\newsec{Quantum field theory}

This section will describe how quantum field theory (QFT)  fits into and provides an example of the UQM framework, and discuss the extra structure imposed by QFT, bearing in mind that we expect to likely need to eliminate some of this structure in a theory describing quantum gravity.

The connection between QFT and the above description of a more general quantum mechanics is most easily understood in Heisenberg picture; here $\calh$ corresponds to the space of Heisenberg-picture states of the field theory.  For a more explicit explanation of the formulation of QFT close to the present picture, see \Wein. One may take different bases of states, corresponding to an ``in" basis, ``out" basis, etc. QFT is typically not explicitly assumed to require an extended descriptor space $\calh^d$.

QFT in flat space  contains states that form irreducible representations of the Poincar\'e group.   
The states include the vacuum, $|0\rangle$, one particle states $|{\vec p},m_I\rangle$ (together with possible spin labels) for the different stable particles $I$, with masses $m_I$, and multiparticle states $|{\vec p_1},{\vec p_2}\rangle$, etc.  The general interacting multiparticle state has a complicated description.  However, a basic tool of QFT is the simplification of such a state that takes place when one describes it in terms of its appearance in the far past or far future.  These ``in" and ``out" states are typically expected to be arbitrarily well-described as free-particle states, described by the momenta of the individual particles, together with their spins.  
In $D$ dimensions the allowed spatial momenta range over ${\bbb R}^{D-1}$. 

Once such a labeling of states is given, much useful data of the theory can be described by giving the S-matrix. This can be defined in terms of these in and out states, as 
\eqn\Sdef{S_{\alpha\beta} = {}_{out}\langle\alpha|\beta\rangle_{in}\ .}
But QFT goes beyond the S-matrix framework, by also assuming that the theory is 
equipped with a hamiltonian $H$ generating unitary time evolution for finite times.

Finally, locality is implemented both in the structure of the multiparticle spectrum, which allows independent particles to be created at different locations on a spacelike slice, and through the structure of the hamiltonian.  In particular, it is typically ensured by assuming the existence of a local hamiltonian density $h(x)$, 
\eqn\hamild{H=\int d^{D-1} x h(x)\ ,}
which commutes with itself at spacelike separations,
\eqn\hamcom{[h(x),h(y)]=0\quad ,\quad (x-y)^2>0\ .}
 Both of these aspects of locality can typically be simply understood through the existence of local fields, such that local gauge-invariant observables constructed from these commute outside the light cone.  (For more discussion, see \Wein, chapter 5.)  These local gauge-invariants can be used both to create particles at distinct locations, and also serve as building blocks for the lagrangian (and thus hamiltonian) density describing local evolution.  Finally, an alternative approach to computing the S-matrix \Sdef\ is to sum over histories, which in the present context are classical field configurations.

\newsec{Nonlocal mechanics about flat space}

We are interested in understanding what parts of the structure of local QFT might need to be dropped in formulating a complete theory incorporating quantum gravity.  While cosmological and other more general contexts are obviously of interest, a simpler case is to consider a theory describing excitations about flat ``Minkowski space" (more specifically, the corresponding vacuum), which we will assume is a solution of quantum gravity.  By this it is meant that there is a quantum state $|0\rangle$ corresponding to the the ``Minkowski vacuum," and there are states describing excitations about this vacuum.
We would like to understand what structure is plausible for such a theory (which will be given the name ``nonlocal mechanics"), and how it would match onto that of quantum field theory in an appropriate limit.

We begin with a description of the states.  We have assumed the existence of the vacuum, $|0\rangle$.  Next, we assume that there are one-particle states.  If $k$  labels momenta of gravitons and $p$ of other matter, these are denoted $|\vec k\rangle$ and $|\vec p,m_I\rangle$ (ignoring spin labels).  Taking as a guide the fact that single particle states with arbitrarily large momenta seem to correspond to a well-defined classical gravitational solution (that of Aichelburg-Sexl\refs{\AiSe}), we will assume that sensible one-particle states exist with arbitrarily high momenta, $\vec k, \vec p \in {\bbb R}^{D-1}$.  Moreover, we also might guess that there is a unitary action of the Lorentz group on such states, acting in the usual way (and preserving $|0\rangle$).

At the two particle level, we expect some modification of this story.  In order to understand this, consider to what extent we expect a precise field-theory construction of arbitrary two-particle in and out states.  For concreteness, let us fix attention on such states at a fixed large negative time $-T$.  Since particles of definite position are rather singular -- having arbitrarily large momenta -- instead consider building up a basis of states from minimum-uncertainty wavepackets.\foot{Indeed, in more precise approaches to quantum field theory, field operators are only defined by integrating against smooth test functions.}  In usual quantum field theory, ignoring gravity, one can construct arbitrary multiparticle states of well-separated particles, with arbitrarily high momenta, and these are very well approximated by free-particle states.  

However, in the presence of gravity, the construction of such field theory states breaks down, at {\it arbitrarily large separation}, for states of sufficiently high energy.  For simplicity, let us illustrate this for states of total momentum zero (i.e. work in the center-of-mass frame).  For two wavepackets at positions $\vec x_1$ and $\vec x_2$, with large $|\vec x_1 -\vec x_2|$, and moderate momenta (and with suitable widths), we have an excellent free-particle description.  However, this description clearly breaks down once the center-of-mass energy, $|p_1+p_2|$ becomes large enough.  This is due to strong gravitational effects, like black holes.  The criterion to avoid such breakdown is (in arbitrary spacetime dimension $D$)
\eqn\locbd{|p_1+p_2|< G|x_1-x_2|^{D-3}\ }
where $G$ is a constant proportional to Newton's constant.
This is known as the {\it two particle locality bound}.  This was stated in \refs{\GiLi}, and more carefully in \refs{\GiddingsBE}.  

Thus  there are expected to be sensible states arbitrarily close to two-particle states (modulo soft-graviton issues, which seem to be a different matter with a more conventional resolution), in the regime where \locbd\ is obeyed, but one loses a free particle description, and quite plausibly a local QFT description, of such states beyond this limit.  This would be in accord with the general {\it nonlocality principle} enunciated in \refs{\GiddingsSJ,\GiddingsBE}, stating that the nonperturbative dynamics which unitarizes gravity is intrinsically nonlocal.  One of the problems of formulating a suitable nonlocal mechanics is that of understanding how to characterize such states in this limit.

These considerations generalize to the case of N-particle states.  Again, imagining a QFT description on a spatial slice at $-T$, multiparticle states with large separations and moderate momenta are very well-described by free-particle states.  But this description breaks down when enough energy is concentrated in a small enough region that gravitational interactions become strong.\foot{It may be that for many purposes there is a good approximate QFT description of the states, as QFT states in the semiclassical geometry of a large black hole, but that this is not a complete and precise description.}
Parallel to \locbd, a criterion for this is the {\it N-particle locality bound}\refs{\LQGST}.  To avoid a breakdown of a local QFT description, one expects to require a condition such as
\eqn\Nlocbd{|\sum_i p_i| < G\, {\rm Max}_{ij} |x_i-x_j|^{D-3}}
for $i,j$ ranging over all subsets of the $N$ particles.  Once this bound is violated, one apparently needs a different description of the states.

One might thus assume that such N-particle states exist and are well-approximated by QFT states in regimes that satisfy the locality bounds \locbd, \Nlocbd.  The proper description of  states outside this regime is a very interesting question.  Given  all such states, an obvious basic physical quantity is the S-matrix, of the form \Sdef.  A reasonable expectation is that the Lorentz group also has a unitary action on such states, and that the S-matrix is Lorentz invariant and unitary.

One also could plausibly expect to have a definition of an asymptotic ``time at infinity," and corresponding hamiltonian/energy for such states.  However, there is not an obvious  fundamental reason to believe that such dynamics is described by a local hamiltonian density as in \hamild, or that there is a precise classical notion of space and time at finite points.

While the S-matrix is a very useful characterization of the physics, one might also expect to have some additional structure that allows approximately local descriptions.  In particular, such descriptions might be expected to match onto the framework of local quantum field theory, in regimes where the latter is valid.  Additional structure of local QFT includes the notion of the state in a region at a given time, where properties of the state are characterized in terms of local operators.
In the context of gravity, one expects such structures to be {\it relational} due to what at low-energies is described as diffeomorphism invariance.  For example, one might find a basis of states that describe the possible appearances of a state at a particular ``time"  (and place) that corresponds to some macroscopic variable of the states which plays the role of a clock.  A suggestive example is the two colliding particles described in section 2.2.  (Such an approximate time variable may -- or may not -- match the  possible asymptotic time.)  Such a description clearly requires a better understanding of the space of states and their unitarily equivalent descriptions.

Beyond this, one might for example wish to recover basic objects of local field theory such as local observables.  A proposal of how these could emerge from more basic operators that are {\it e.g.} diffeomorphism-invariant in the low-energy setting is  via the relational observables of \refs{\Dewitt,\GMH}.  Specifically, we expect that in certain states, certain operators approximately reduce to the local observables of QFT.  If one is describing these operators in a limit where QFT and a metric are good approximate notions, the simplest such operators take the form 
\eqn\prot{\calo = \int d^Dx \sqrt{-g} {\hat \calo}(x)\ }
where ${\hat \calo}(x)$ is a local operator.  Such a construction ensures that the operator $\calo$ respects the symmetries of the low-energy theory, which are diffeomorphisms.  Examples of how such ``proto-local" observables can, in appropriate states, reduce to local observables are given in \refs{\GMH,\GaGi}.  One may, in turn, wish to characterize states by the action of such observables, or of more general operators.

\newsec{Aspects of nonlocal mechanics for cosmology}

Such a quantum-mechanical framework should also apply to more general contexts, and in particular cosmology.  The framework described here appears to have the basic ingredients to do so.

Specifically, the space $\calh$ can be taken to comprise different quantum states of a cosmology (or of ``the universe").  Basic quantities are then the overlaps $\langle \psi'|\psi\rangle$ between different states.  However, to provide an interpretational context for these quantities, one needs further knowledge about the labels on the different possible states.
In the special circumstances where this cosmology has a regime well-described by semiclassical physics, one should be able to parametrize the possible states in terms of their appearance as  states of a QFT in a given semiclassical background, with a time variable given by a semiclassical feature or features of the state.  

For example, let  $|\psi\rangle$ be an allowed state of the cosmology.\foot{In the spirit of \HartleAI, there could be a unique such state.}  In the case where there are states well-described by semiclassical physics, we may have a set of states  (or more generally, descriptor states) $|\psi_I\rangle$ where $I$ might be labels well-approximated by those of QFT states in a background.  Or, going one step further, there may be a set of such states parametrized by a dynamical variable that acts as a semiclassical time $t$; thus one might have states of the form $|\psi_I, t\rangle$ for each value of $t$ in some range.  Then we could say that the amplitude for the state $|\psi\rangle$ to appear as such a state is
\eqn\semiamp{\langle \psi_I, t\ |\psi\rangle\ .}

Likewise, the states may also be characterized by how certain operators act on them.  There may be a special class of such operators, that approximately reduce to local operators of QFT (or products thereof), along the lines of the proto-local operators (or their generalizations) described above and in \GMH.  Alternatively, in cases where an appropriate decoherence functional is defined, it could produce the amplitude for a particular sequence of alternatives; these give us probabilities in cases where the alternatives decohere.

As an example consider  de Sitter space, which is discussed from a related viewpoint in \refs{\GiMa}.  Here there is a space of states that we expect to be well-described in perturbative gravity.  These are the states satisfying the {\it de Sitter locality bound}\GiMa.  This is an analog of the bound \Nlocbd, and takes the form
\eqn\dslocbd{ F < G R^{D-3}}
where
\eqn\fluxdef{F = \int d^{D-1}x\sqrt{g_{D-1}} T_{00}}
is the flux of energy through the ``neck" of de Sitter space, and $R$ is the de Sitter radius.  In addition, we might expect that there is a nonperturbative completion of the space of states, so that the complete space of states $\calh$ describing the nonlocal mechanics of global de Sitter space has dimension $\exp\{S_{dS}\}$.  The physics would then be described by overlaps of the different states, and/or by the properties of relational observables acting on the states.  We may or may not find that the asymptotics are such that there is in fact a unitary S-matrix.
Among the states are those that have features that evolve semiclassically; for those states, we can give an approximate description in terms of QFT variables evolving with respect to the space and time variables defined by the semiclassical features.

This example also illustrates the possible utility of going beyond the space $\calh$ to a larger descriptor space $\calh^d$.  Specifically, one may wish to give the amplitude for an allowed state $|\psi\rangle\in \calh$ to look like a state $|\varphi\rangle_d$ that can for example be described as a specific quantum field theory state at some large radius of the universe.  The latter state may {\it not} have any sensible description as a state of global dS space, as it for example may, if  ``evolved back," have a flux $F$ that is too large, or otherwise violate the locality bound \Nlocbd.  Thus it may not lie in the space $\calh$ of physical states.  Nonetheless, such states may be quite useful in describing the different possible appearances of an allowed state $|\psi\rangle$, through the quantities ${}_d\langle \varphi | \psi\rangle$, or through matrix elements of relational observables.

\newsec{Further discussion}

The framework of universal quantum mechanics described in this paper is simply a proposed broad quantum mechanical and interpretational framework for a theory of quantum gravity that does not have some of the standard properties of local field theory.  In particular, as the above discussion has illustrated, some of the features of ordinary (or generalized) quantum mechanics are {\it not} generically  expected to be part of such an interpretational framework, so the current proposal works towards a minimal set of postulates.

Of course, much more structure is needed in order to formulate a complete theory incorporating quantum gravity.  A central point, though, is that structures such as temporal and spatial location are expected to emerge from the dynamics of the theory, in certain states, and thus not be general features of the broader framework.  No fundamental spacetime is assumed; in general it might emerge as an approximate concept.  Since such a dynamical theory is in particular not expected to have the specific property of locality, which is crucial to quantum field theory, we have used the name ``nonlocal mechanics" for such a theory.   

The framework described represents a generalization of the old S-matrix program (see, {\it e.g.}, \refs{\Eden}) in different directions.  While postulates one and two correspond to postulates of that program, one does not take as fundamental postulates the existence of only short range interactions or the usual assumptions of fundamental locality/causality.  In the general context, Lorentz invariance is also not part of the axioms.  Moreover, in the cosmological context one needs to talk about states without use of asymptotically flat regions.  This is expected to be accomplished by states that are labeled relationally, as opposed to simple in and out states.

A crucial question is to determine what symmetry and other dynamical principles should govern a correct gravitational theory within this framework.  Of course, it is conceivable that the relevant theory arises from some preexisting structure such as string theory.  In weak-coupling regimes, string theory appears to furnish an S-matrix for gravitational scattering, and it is hoped that with further developments it will also do so at strong coupling and supply similar constructions for cosmology.  One might also anticipate the possibility of constructing suitable relational observables in string theory, as suggested for example in \GaGi, that approximately reduce to the local observables of QFT.  However, without knowing that string theory has sufficient structure to supply the needed physical constructions and is the correct theory, it is perhaps best  
not to prejudice the issue, and to pursue the relevant principles of nonlocal mechanics in whatever theoretical form they appear.

One important and likely strong constraint is that nonlocal mechanics should reduce to local quantum field theory plus semiclassical gravity in appropriate limits.  The locality bounds \locbd, \Nlocbd, and \dslocbd\ are suggested parametrizations of part of the boundary between the domains of fully nonlocal mechanics and local QFT.  Finding such a nonlocal and quantum mechanical structure in which QFT  approximately embeds in this fashion seems like a very difficult order to fill, but that may be good news as it suggests strong constraints on the problem that could yield important guidance towards its solution.

\bigskip\bigskip\centerline{{\bf Acknowledgments}}\nobreak

I wish to thank D. Marolf, R. Porto, M. Srednicki, and especially J. Hartle for 
 valuable conversations and for comments on earlier drafts.
This work   was supported in part by the Department of Energy under Contract DE-FG02-91ER40618,  and by grant  RFPI-06-18 from the Foundational Questions Institute (fqxi.org).

\listrefs
\end